\documentclass[useAMS,usenatbib]{mn2e}
\usepackage{graphicx}

\newcommand{\HI}{\hbox{{\rm H}{\sc \,i}}}

\newcommand{\lya}{\hbox{{\rm Ly}$\alpha$}}

\newcommand{\Ha}{\hbox{{\rm H}$\alpha$}}

\newcommand{\msun}{\hbox{M$_{\odot}$}}
\newcommand{\zsun}{\hbox{Z$_{\odot}$}}
 \newcommand{\rhosun}{\hbox{M$_{\odot}$\, Mpc$^{-3}$}}
\newcommand{\cmsq}{\hbox{cm$^{-2}$}}

\newcommand{\hMpcc}{\hbox{$h_{100}^{3}$~Mpc$^{-3}$}}

\newcommand{\kms}{\hbox{${\rm km\,s}^{-1}$}}

\def\bsp_small{\vspace{0.5cm}\small\noindent This paper
has been typeset from a \TeX / \LaTeX\ file prepared by the author.}

\title[The metals in BX galaxies]{The missing metals problem: II. How
many metals are in $z\simeq2.2$ galaxies?}

\author[N. Bouch\'e, M. D. Lehnert, C. P\'eroux]{Nicolas
Bouch\'e$^1$\thanks{E-mail: nbouche@mpe.mpg.de (NB)}, Matthew
D. Lehnert$^1$, C\'eline P\'eroux$^2$\\ $^1$Max Plank Institut f\"ur
extraterrestrische Physik, Giessenbachstra\ss e, D-85748 Garching,
Germany \\ $^2$European Southern   Observatory, Karl-Schwarzschild-Str 2,
D-85748 Garching, Germany}

\begin{document}

\date{Accepted ---.
      Received ---;
      in original form ---}

\pagerange{\pageref{firstpage}--\pageref{lastpage}}

\maketitle

\label{firstpage}

\begin{abstract}
In the context of the ``missing metals problem'', the contributions of the
UV-selected  $z\simeq2.2$ ``BX'' galaxies and $z\simeq2.5$ ``distant red
galaxies'' (DRGs) have not been discussed previously.  Here we show that:
(i) DRGs only make a marginal contribution to the metal budget ($\sim
5$\%); (ii) BX galaxies contribute as much as 18\%\ to the metal budget;
and (iii) the $K$-bright subsample ($K<20$) of the BX sample (roughly
equivalent to the `BzK' selected samples) contributes roughly half of this
18\%, owing both to their larger stellar masses and higher metallicities,
implying that the rare $K$-bright galaxies at $z>2$ are a major source of
metals in the budget.  We showed in the first paper of this series that
submm galaxies (SMGs) brighter than 3~mJy contribute $\sim$5\%\ ($\la9$\%
as an upper limit) to the metal budget.  Adding the contribution of SMGs
and damped Ly$\alpha$ absorbers, to the contribution of UV selected
galaxies, implies that at least 30\%\ of the metals ({\it in galaxies})
have been accounted for at $z\simeq2$.  The cosmic metal density thus accounted for is
$\rho_{Z,\rm galaxies}\simeq 1.3\times 10^6 \;\rhosun$ or in terms of
the closure density, $\Omega_Z=9.6\times10^{-6}$.  This is a lower limit
given that galaxies on the faint-end of the luminosity function are not
included.  An estimate of the distribution of metals in local galaxies as
a function luminosity suggests that galaxies with luminosity $<$L$^{\star}$
contribute about half of the total mass of metals.  If the metals
in galaxies at z$\sim$2 are similarly distributed then faint galaxies
alone cannot solve the `missing metals problem.'  Galaxy populations
at z$\sim$2 only account for about 50\% of the total metals predicted.
\end{abstract}



\section{Introduction}

For a given initial mass function (IMF), the total expected
amount of metals $\rho_{Z,\rm expected}$ formed at a
given time $t$ is simply the integral of the star formation
history (SFH) or star formation rate density \citep[SFRD;][and
others]{LillyS_96a,MadauP_96a,GiavaliscoM_04a,HopkinsA_04a}: i.e.,
$\rho_{Z,\rm expected} = \dot{\rho}_{\star}(t) \times <p_z>$,
where $<p_z>$ is the mean stellar yield \citep{SongailaA_90a}.
\citet{MadauP_96a} found that $<p_z>=\frac{1}{42}$ or 2.4\%\ using a
Salpeter IMF extending from 0.1 to 125~\msun~\footnote{\citet{MadauP_96a} notes
that the conversion is fairly insensitive to the IMF as long as the stellar population
 extends as a power law to massive stars, 50-100~\msun.} and the type II stellar yields (for solar metallicity)
$p_z(m)$ from \citet{WoosleyS_95a}.  After integrating the SFH over the
redshift $z$ range from 4 to 2 (or 1.75~$h_{70}^{-1}$~Gyr), we find that
the total co-moving metal density is:
\begin{eqnarray}
\rho_{Z,\rm expected}&=&4.0\times 10^6\; \rhosun\label{eq:metals:madau}  ,
\end{eqnarray}
\citep{BoucheN_05e,PettiniM_03b} using the SFH parameterized (in a LCDM cosmology)
either as in \citet{ColeS_01a} or by a constant star formation rate
(SFR) beyond $z=2$. Equation~\ref{eq:metals:madau} is about 25\%\ of the
$z=0$ metals.

At redshifts $z\simeq2-3$, our knowledge of the cosmic metal
budget is still highly incomplete, contrary to the situation at redshift $z=0$
\citep[e.g.][]{FukugitaM_04a}.  Until very recently, it was
thought that only a small fraction (20\%) of the budget is
actually accounted for when one adds the contribution of the \lya\
forest ($N_{\HI}=10^{13-17}$~\cmsq), damped \lya\  absorbers (DLAs)
($N_{\HI}>10^{20.3}$~\cmsq), and  galaxies such as Lyman break galaxies
(LBGs) \citep{PettiniM_99a,PagelB_02a,PettiniM_03b,WolfeA_03b}.

Over the last decade, numerous samples of $z>2$ galaxies have emerged thanks to
the increasing size and sensitivity of near-infrared detectors and
to the availability of ultraviolet-sensitive instruments. On the
one hand, the $z=3$ Lyman break technique \citep[e.g.][]{SteidelC_99a}
was extended  to lower redshifts ($z\sim1.5$--2.5) using different
$U_nG\cal R$ colour criteria \citep{SteidelC_04a}.  Samples selected in
this way are referred to as  `BX/BM' galaxies \citep{AdelbergerK_04a}.
Using near-infrared imaging to select red galaxies with $J-K_s>2.3$
colour, the Faint Infrared Extragalactic Survey (FIRES)   \citep{FranxM_03a,vanDokkumP_03a} unveiled
significant numbers of galaxies at $z\sim2.5$ which they dubbed
`distant red galaxies' (DRGs).  The FIRES near-infrared selection
includes both passively evolving (PE) and reddened star-forming (SF)
galaxies with $E(B-V)>0.3$ \citep{ForsterSchreiberN_04a}.  
The $K<20$ criterion (hereafter `K20'; Cimatti et al. 2002)
and in particular the $B-z$ vs. $z-K$ colour criteria (hereafter `BzK'; Daddi et al. [2004,2005])
also revealed a significant number
of galaxies with $1.5<z<2.5$ with a range of star-formation histories
including both SF and PE galaxies.

This paper is the second in our series to estimate the contribution
of various galaxy populations at $z\sim2$ to the total metal  budget.
In \citet{BoucheN_05c} (paper~I), we showed  from  current observations
of $z\simeq 2$ submm galaxies (SMGs) that SMGs (brighter than 3~mJy)
contain $\sim$5\%\ (and certainly $\la$9\%) of the metal budget at
$z\simeq 2$.  In this paper, we discuss the contribution of $z\simeq2.2$
colour selected galaxies to the metal budget.  We will show that  colour
selected galaxies (`BX' galaxies) contribute significantly ($\sim18$\%)
to the metal budget.  The $K$-bright ($K_s<20$, hereafter BX$+$K20)
subsample contributes 8\%\ to the metal budget, or almost half of the
18\%.  Since these samples are all magnitude limited, we discuss likely
contribution of the sample of galaxies at z$\sim$2 if one considers the
faint end of the luminosity function using the metal-luminosity relation
in local galaxies.  Such an analysis suggests that galaxy populations
at z$\sim$2 only contribute about 50\% to the total metal budget.
In paper~III, we will  explore whether the remaining metals have been
expelled from small galaxies into the intergalactic medium (IGM).

In the remainder of this paper, we used $\Omega_M=0.3$,
$\Omega_\Lambda=0.7$, $H_o=70~h_{70}$~\kms~Mpc${-1}$ and $h=h_{100}$.

\section{The metal budget at $z\sim$2.2}

\subsection{The contribution of `BX' galaxies}
\label{section:BX:observations}


Using colour selection criteria similar to those of
$z=3$ LBGs, Steidel and collaborators have constructed large samples of galaxies
at $z\simeq 2.2$ (BX) and 1.7(BM).  The  median redshifts of these
samples are $2.22\pm0.34$ (BX), and $1.70\pm0.34$(BM), respectively
\citep[e.g.][]{AdelbergerK_04a}.


The observed number density of BX and BM galaxies ($n_{\rm
BX/BM}$) is somewhat higher than $z=3$ LBGs.  Indeed, according to
\citet{AdelbergerK_05a}, $n_{\rm BX}=(6\pm3)\times 10^{-3}$~\hMpcc,
and $n_{\rm BM}=(5\pm2.5)\times10^{-3}$~\hMpcc, whereas $n_{\rm
LBG}=(4\pm2)\times10^{-3}$~\hMpcc. We summarize these numbers in
Table~\ref{table:summary} for $h_{70}$.


\citet{ShapleyA_05a} derived the stellar masses ($M_\star$) of 72 spectroscopically-confirmed
$z=2.2\pm0.3$ galaxies using UV to mid-IR observations. The mean $M_\star$
of their sample is  $<\log M_\star>=10.3\pm0.5$ for a Salpeter IMF extending from 0.1 to 100 \msun.


The last piece of information needed to estimate the contribution of
the BX galaxies to the metal budget is their metallicity.  Few BXs
have had their gas phase metallicities published. For instance, \citet{ShapleyA_04a} found that the
metallicities of 7 BX galaxies are close to solar, but the
median of a much larger sample appears to be about half solar \citep{PettiniM_05a}.

Using these estimates, we find that BX galaxies contribute:

\begin{equation}
\rho_{Z,\rm BX}\simeq 3.8\times10^5\;    \frac{<Z>}{0.5 \,\zsun}\;   \rhosun
\end{equation}
to the cosmic metal density or about 10\%\ of the expected metal
density at z$\sim$2 (equation~\ref{eq:metals:madau}).

BX galaxies with K-band magnitudes brighter than 20 (hereafter BX$+$K20)
appear to be more evolved systems.  Indeed, they seem to have higher
stellar masses [$<\log M_\star>=11\pm0.6$, \citet{ShapleyA_05a}],
and are more metal rich: their median metallicity is slightly below
solar [$Z\sim 4/5 Z_\odot$, \citet{PettiniM_05a}].  
However, they are significantly
rarer: their number density is a factor of 10 lower 
[$n\sim 1.7\times10^{-4}\;h_{70}^3$~Mpc$^{-3}$, \citet{ShapleyA_05a}]. 
 The outcome is that   $K_s$-bright BX galaxies contribute significantly to the metal
budget:

\begin{equation}
\rho_{Z,\rm BX,K<20}\simeq 3.0\times10^5\;    \frac{<Z>}{0.8\,\zsun}\;   \rhosun
\end{equation}

or 8\% of the estimated total metal density (equation~\ref{eq:metals:madau}).
These numbers are summarized in Table~\ref{table:summary}.

Unlike for the SMGs in paperI, we have not applied a correction for the
``duty cycle'', the fraction of cosmic time over which they meet the
observational selection criteria.  The duty cycle of BXs can be estimated
from the  comparison of  the clustering strength to the observed number
density \citep{AdelbergerK_05a}.  \citet{AdelbergerK_05a} find that
the duty cycle for the ``BM/BX'' galaxies is $\sim$1.0, suggesting no
correction to our estimate is required.

\subsection{The contribution of DRGs}
\label{section:DRgs:observations}

Red near-infrared colour selection of galaxies
\citep[such as DRGs with $J-K_s>2.3$][]{FranxM_03a,vanDokkumP_03a}   was designed to
discover large numbers of $z\sim2.5$ evolved galaxies with a strong Balmer break.
DRGs are massive galaxies and include both heavily reddened  SF  galaxies,
and PE galaxies   \citep{ForsterSchreiberN_04a,LabbeI_05a}.



The space density of DRGs is about $1\times10^{-4}\;h_{70}^3$~Mpc
\citep{LabbeI_05a,ReddyN_05a}.  \citet{ForsterSchreiberN_04a} found,
from modelling the spectral energy distributions (SEDs) of  DRGs, that their
mean stellar mass is about 5 times higher than z$\sim3$ LBGs, or $M_\star
\simeq 1\times 10^{11}$~\msun (for a Salpeter IMF between 0.1 and 100
\msun) and only varies by a factor of $\sim$2 depending on the star
formation history assumed.


\citet{vanDokkumP_04a} estimated the metallicity of a couple of DRGs from Keck
NIRSPEC spectra.  They found that the metallicities, as determined using
$[N II]/\Ha$, are high, 1-1.5 times the solar value.  From the long
slit kinematics, \citet{vanDokkumP_04a} estimated that the dynamical
masses are about $\sim1\times10^{11}$~\msun, in agreement with the mass
estimates using the SED modelling of \citet{ForsterSchreiberN_04a}.


Based on these published results, we find that the cosmic metal
density contributed by DRGs is:

\begin{equation}
\rho_{Z,\rm DRGs}\simeq 2\times10^5\;    \frac{<Z>}{1\,\zsun}\;   \rhosun \label{eq:metals:drgs}
\end{equation}

or 5\%\ of the expected metal density (equation~\ref{eq:metals:madau}).

\section{Discussion \& Conclusions}

Table~\ref{table:summary} summarizes the observations discussed here.
The properties of $z=3$ LBGs are listed for comparison.

\begin{table*}
\caption{Summary of properites of $z>2$ galaxies.\label{table:summary} References: 
(a)  \citet{AdelbergerK_05a},
(b) \citet{PettiniM_05a},
(c) \citet{ShapleyA_05a},
(d) \citet{PettiniM_01a}
(e) \citet{ShapleyA_04a}, 
(f) \citet{ReddyN_05a},
(g) \citet{LabbeI_05a},
(h)  \citet{ForsterSchreiberN_04a},
(i) \citet{vanDokkumP_04a}.
}
\begin{tabular}{lllllll}
	&  $<z>	$	&	$n$ 	($h_{70}^{3}$~Mpc$^{-3}$)       &   $Z/Z_\odot$  	
	&  $M_\star$ (\msun)		  	& $\rho_Z$ ($M_\odot$~Mpc$^{-3}$) & $\Omega_Z$  \\
	&		& $ 10^{-3}$ &		&   $ 10^{10}$      & $10^5$ & $10^{-6}$\\
	\hline		 									     
BX	&$2.2\pm0.3$	&  $2\pm1$$^{\rm a}$	&  0.5~$^{\rm b}$		&  2~$^{\rm c}$	&    	3.8 & 2.8  \\
BX$+$K20&$2.2\pm0.3$	&  $0.2$~$^{\rm c}$		&   0.8~$^{\rm b}$	&  10~$^{\rm c}$&    	3.0 & 2.2  \\
DRGs	&$2.5\pm0.4$	&  $0.1$~$^{\rm f,\rm g}$	&   1.0~$^{\rm i}$ 	&  10~$^{\rm h}$&    	1.9 & 1.4  \\
\hline
LBGs	&$3.0\pm0.3$	&  $1.3\pm0.7$~$^{\rm a}$	&   0.33~$^{\rm d}$	&  1.2~$^{\rm e}$&    	3.0  & 2.2  \\
\hline
\end{tabular}
\end{table*}

From the observations discussed above, we find that the metal density
in BX galaxies is:
\begin{equation} \rho_{Z,\rm BX}=3.8\times 10^5\;\frac{<Z>}{0.5 Z_\odot}\rhosun\;, \label{eq:metals:bx:all}
\end{equation} 
or 10\%\ of the metal budget, assuming a mean metallicity $<Z>$ of
one-half solar and the above co-moving space densities.

The more evolved  sub-sample of the BX sources, namely BX$+$K20 galaxies,
are rarer (by a factor of 10), but more massive (by a factor of 5),
and more metal rich (by a factor of 2).  Thus, they double the total
contribution of `BX' galaxies to the metal budget.  The metal density
of this subsample is:
\begin{equation}
\rho_{Z,{\rm BX+K20}}=3.0\times 10^5\;\frac{<Z>}{4/5 Z_\odot} \rhosun \label{eq:metals:bx:k20}
\end{equation} 
or 8\%\ of the metal budget.  Thus, the rare $K$-bright galaxies at
$z>2$ are a major source of metals in the budget.  We are still far
from closing the metal budget however.  If we take these numbers  at
face-value, as much as 18\%\ of the metals could be in BX galaxies.


Regarding the `BzK' selection of $z\sim2$ galaxies by \citet{DaddiE_04a},
we note that the mean redshifts for BzK/SF (star-forming) galaxies
and BzK/PE (passive evolving) galaxies is $<z>\simeq2\pm0.3$ and $1.7\pm0.2$,
respectively \citep{ReddyN_05a}. Focusing on the $z>2$ subsample (i.e.,
BzK/SF), \citet{ReddyN_05a} showed that the `BX' selection criteria
and that the `BzK/SF' selection criteria overlap significantly ($\sim 80$\%)
to $K_s=22$.  We therefore view these two populations (BX and BzK/SF) as
two strongly overlapping samples which do not require separate analyses
of their contribution to the metal budget at z$\sim$2.

Having now considered a substantial number of samples of galaxies selected
at z$\sim$2, we can now estimate the total metal budget in galaxies. We
have estimated that 5\% ($\la 9$\%) of the metals are in $z\sim 2.4$ SMGs
(brighter than 3~mJy; as shown in paper~I), 18$+$5\%\ in UV-selected and
$J-K$-selected galaxies, and  5\%\ in DLAs \citep[][and the forthcoming
paper~III]{PettiniM_03a}.  The resulting sum is:
\begin{equation}
\rho_{Z,\rm galaxies}\simeq 1.3\times 10^6 \;\rhosun\label{eq:total:galaxies}
\end{equation}
or $\sim$33\%\ of the expected metal density at $z\simeq2$ (equation~\ref{eq:metals:madau})~\footnote{We
 note that at $z\sim3$, the ``missing metals
problem'' likely remains more acute. Unlike the ``BM/BX'' galaxies,
LBGs only contribute about 5-10\%\ of the total metals at z$\sim$3.}.

However, equation~\ref{eq:total:galaxies} is a lower limit given that we really only considered
the bright end ($\ga$L$_{\star}$) of the galaxy luminosity function.
One obvious criticism of our approach is that given the limited depth of
the selection images and the obvious incompleteness, could the metals
hide in low luminosity/low mass galaxies?  While it is difficult to
estimate what the contribution of the faint end of the luminosity
function might be to the total metal budget, perhaps considering the
local population of galaxies might be useful as a guide.  In the local
Universe, while galaxies at the faint-end of the luminosity function are
more numerous than the luminous galaxies, they are also more metal-poor
\citep[e.g.,][]{TremontiC_04a}.

\begin{figure}
\centerline{\includegraphics[width=75mm]{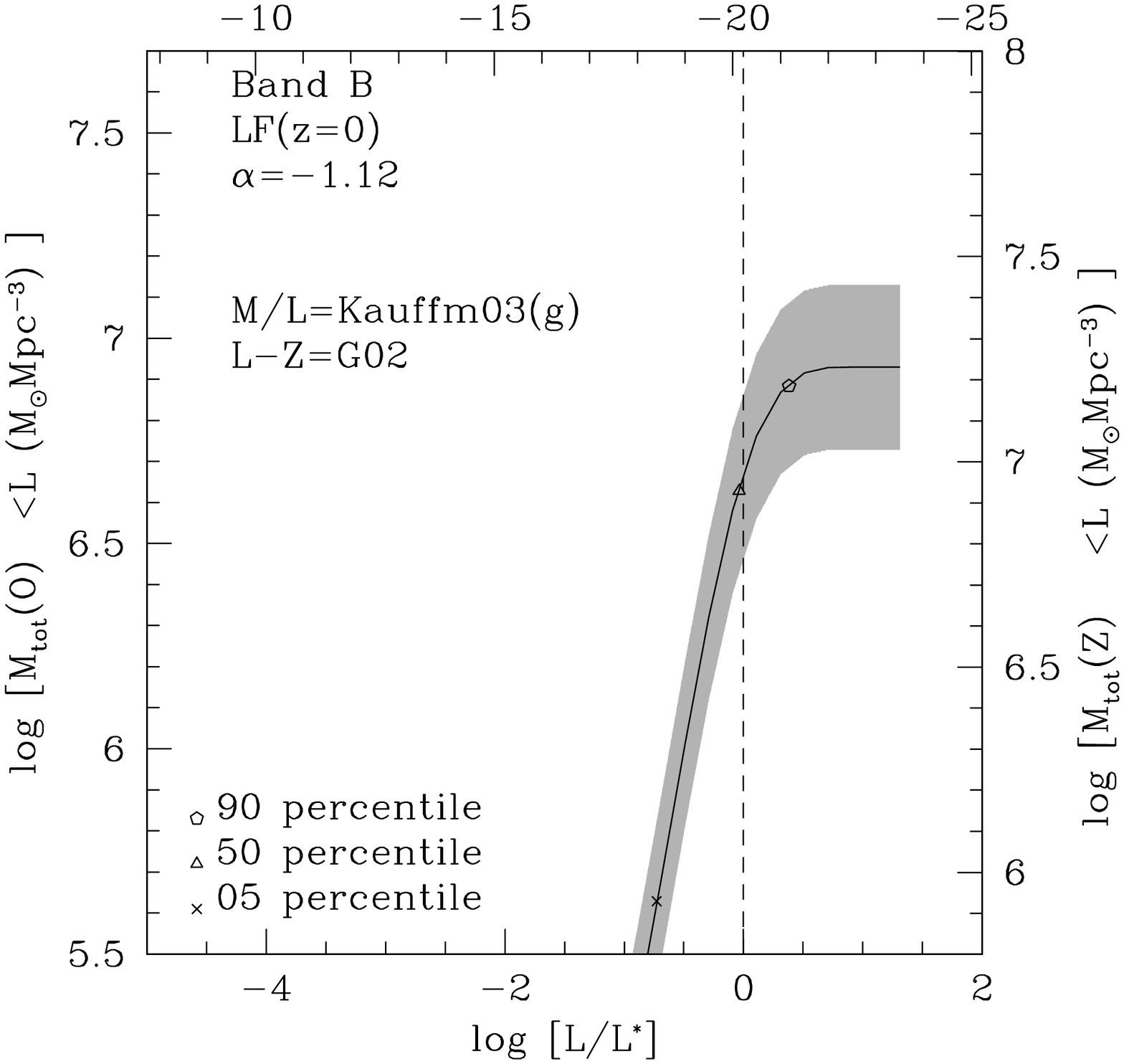} }
\caption{The cumulative Oxygen  cosmic density using the $z=0$ luminosity function from
\citet{DriverS_05a}, the LZ relation from \citet{GarnettD_02a} and the
M/L from \citet{KauffmannG_03a}. The right-axis shows the cosmic metal
density.  The 5th, 50th  and 90th percentiles are shown as a cross,
triangle and circle respectively.  Because of the very opposite slopes
of the LF and of the LZ relation, the 50th percentile arises at around
$L\simeq L_\star$. As a consequence, 50\%\ (or more) of the metal budget
lies in the faint end of the LF.  Note that the metal density reaches
$\sim2\times10^7$~\rhosun, close to the metal density obtained from
integrating the SFH, i.e. $1.8\times10^7$~\rhosun, showing that the
normalization appears to be correct.}
\label{fig:extrapol}
\end{figure}

For the local Universe, we estimate the contribution
of the faint-end of the luminosity function   $\phi(L)$
using the well-known luminosity-metallicity ($LZ$) relation
\citep[e.g.][]{TremontiC_04a,GarnettD_02a}.  Using mass-to-light
ratios [$M/L(L)$] from \citet{KauffmannG_03a}, the $z=0$ $LZ$ relation
($O/H(M)$ or $Z(M)$) from \citet{GarnettD_02a} and $\phi(L)$ from
\citet{DriverS_05a}, one can estimate the cumulative metal density
($\int \phi(L)M/L(L)Z(M)$).  Fig.~\ref{fig:extrapol} shows the
cumulative Oxygen density $M(O)$ at $z=0$.  The right axis shows
the total metal density $M(Z)$.  The 5th, 50th  and 90th percentiles
are shown.  From Fig.~\ref{fig:extrapol}, one sees that, at $z=0$,
50\%\ of the metals are in galaxies fainter than $L_\star$.
Note that the metal density reaches $\sim2\times10^7$~\rhosun,
close to the metal density obtained from integrating the SFH down to
$z=0$, i.e. $\rho_{Z,z=0}=1.8\times10^7$~\rhosun, showing that the
normalization must be approximately correct.  Constraints on the $LZ$
relation at $z>2$ are just becoming available \citep[][C.C. Steidel,
private communication]{SavaglioS_05a,ErbD_05a}.  

Based on these local results, at least 50\%\ of the metals in
galaxies are in objects on the faint-end of the luminosity function
(L$\la$L$^{\star}$).  The limiting depths of the surveys that find
z$\sim$2 galaxies reach to or even beyond L$^{\star}$.  If we assume for
simplicity that our numbers and estimates are appropriate for L$^{\star}$
and more luminous galaxies and that the z$\sim$2 galaxies have a similar
relative fraction of metals above and below L$^{\star}$, it would suggest that we
need to correct our total metal estimate by about a factor of 2. This is
likely an over-estimate of the true correction necessary for the faint end
of the luminosity function and shows that the total contribution to the
metal budget is roughly 60\%\ or less.  In addition, this sum might also
count   some of the galaxy types twice, given that we effectively extrapolate
to the faint end of the LF for each of the samples.  Given these caveats,
one can easily account for $\sim 30$---60\%\ of the metal budget. We will
discuss in paper~III the possibility that the remaining 
missing metals could have been ejected from small galaxies via
galactic outflows into the IGM.

\section*{Acknowledgments}
We thank the referee for a prompt report;  M. Pettini for interesting discussions and for sharing his
results before publication; and N. F\"{o}rster Schreiber for sharing her
extensive knowledge of DRGs.


\bsp_small

\label{lastpage}

\end{document}